\documentclass{article}
\usepackage{graphicx}
\usepackage{hyperref}
\usepackage{authblk}
\date{ }
\begin{document}

\title{Verlinde gravity effects on the orbits of the planets and the Moon in the Solar System}
\author[1]{Youngsub Yoon}
\author[2,3]{Luciano Ariel Darriba}

\affil[1]{\it{Dunsan-ro 201, Seo-gu}\\\it{Daejeon, 35245 Korea}}
\affil[2]{\it{Instituto de Astrof\'{\i}sica de La Plata, CCT La Plata-CONICET-UNLP\\
  Paseo del Bosque S/N (1900), La Plata, Argentina.}}
\affil[3]{\it{Facultad de Ciencias Astron\'omicas y Geof\'\i sicas, Universidad Nacional de La Plata.\\
  Paseo del Bosque S/N (1900), La Plata, Argentina.}}
\renewcommand\Affilfont{\itshape\small}

\maketitle

\begin{abstract}
In this work, we address the effects of a phenomenon known as Verlinde gravity. Here we show that its effect over the planets and the Moon in our solar system is quite negligible. We find that the Verlinde gravity effects on the orbits of planets are at least 10 times smaller than the precision with which we can determine the Sun's mass, and the one on the orbit of the Moon is about 100 times smaller than the precision with which we can determine the Earth's mass. These results let us infer that statements in the literature that Verlinde gravity is ruled out by the observed motion of planets in our solar system aren't correct.
\end{abstract}

\section{Introduction}
In 2010, Verlinde proposed what he called ``entropic gravity'' or ``emergent gravity''. This concept states that gravity is an emergent phenomenon due to the entropy change as objects change their distance \cite{entropic}. It attracted considerable attention in theoretical particle physics community, when the work was first published. He showed how Newton's universal law of gravitation and general relativity can be derived from entropic gravity. In 2016, Verlinde kept developing his theory further and successfully explained Tully-Fisher relation \cite{Verlinde}. In other words, his new theory made different predictions than the Newtonian gravity or general relativity. The strict area law for entanglement entropy implies general relativity. However, he suggested that there is a non-zero volume law contribution to the entanglement entropy due to the thermal excitations responsible for the de Sitter entropy. These excitations constitute the positive dark energy that accelerates our Universe, which is similar to de Sitter space. However, his equations are not immediately applicable to more wide range of phenomena, as they are only valid in the presence of spherical symmetry. Thus, in another paper \cite{VerlindeYoon} (in preparation), one of us obtained an expression that is valid in general cases that have no spherical symmetry. We use that expression in this work to show that some claims stating that Verlinde gravity is ruled out by the observed motion of the planets in our solar system are false. 

The organization of this paper is as follows. In Section 2, we explain Verlinde gravity. In Section 3, we discuss the strategy developed to calculate the Verlinde gravity effects on the planets in the subsequent sections. In Section 4, we calculate how the Verlinde gravity affects the gravitational forces exerted on the planets. In particular, we will see that the gravitational force on a planet increases by a factor $(1+\lambda)$, where $\lambda$ is a constant that differs from planet to planet. Therefore, a non-zero $\lambda$ can be interpreted as the difference between the gravitational mass and the inertial mass. We will also see that the value of $\lambda$ depends on the radial density profile of a planet. In Section 5, we will calculate $\lambda$ for each planet. In Section 6, we will compare the orbit of planets with $\lambda$ obtained in Section 5. In Section 7, we present our conclusions.

\section{Verlinde gravity}
Verlinde gravity \cite{Verlinde, VerlindeYoon} is given by
\begin{equation}
g=\sqrt{g_B^2+g_D^2},\label{gB2gD2}
\end{equation}
where $g=|\vec g|$ is the total gravity, $g_B=|\vec g_B|$ is the gravity due to the baryonic (i.e., visible) matter, and $g_D=|\vec g_D|$ is the gravity due to the apparent dark matter. From the form of the above equation, it is tempting to believe that $\vec g_B$ and $\vec g_D$ are perpendicular. However, the direction of $\vec g$ is the same as the directions of $\vec g_B$ and $\vec g_D$. ($\vec g_B$ and $\vec g_D$ are parallel.) Let's give a brief interpretation of \ref{gB2gD2}. The gravitational energy is proportional to $g^2$, and the above equation says that the total gravitational energy is the sum of the gravitational energy due to the visible matter and the one due to the apparent dark matter \cite{VerlindeYoon}. Verlinde constructed Verlinde gravity by an analogy with elastic matter, and one can interprete the gravitational energy as the elastic energy of strained matter. 

In Eq. (6.11) of \cite{Verlinde}, Verlinde relates what he calls $\Sigma_{Di}$ (``surface mass density vector'' of apparent dark matter) with $g_{Di}$ (the gravity due to the ``apparent'' dark matter) as follows: 
\begin{equation}
\Sigma_{Di}=-\left(\frac{d-2}{d-3}\right)\frac{g_{Di}}{8\pi G},
\end{equation}
where the subscripts $i$ denote the $i$th vector component. He noted that one could avoid annoying factors by working with $\Sigma_D$ instead of $g_D$. But here, we need to restore them. Plugging $d=4$, we get
\begin{equation}
\Sigma_{D_i}=-\frac{2g_{Di}}{8\pi G}\label{SigmaDi}.
\end{equation}
In Eq. (7.37) of \cite{Verlinde}, Verlinde relates $\Sigma_D$ with the baryonic variables as follows:
\begin{equation}
\left(\frac{8\pi G}{a_0}\Sigma_D\right)^2=\left(\frac{d-2}{d-1}\right)\nabla_i\left(\frac{\Phi_B}{a_0}n_i\right),
\end{equation}
where 
\begin{equation}
a_0=cH_0,\qquad\vec n=\frac{\vec g_B}{g_B}\label{vecn},
\end{equation}
being $H_0$ Hubble's constant, $n_i(\vec n)$ the direction of gravity, and $\Phi_B$ is a potential similar to the Newtonian potential; while the Newtonian potential is related to the time time component of the metric, $\Phi_B$ is related to the space space component of the metric. In \cite{VerlindeYoon}, one of us showed that
\begin{equation}
\Phi_B=\frac{2g_B}{\nabla_i n_i}\label{nablaini}.
\end{equation}
As an aside, in the presence of spherical symmetry, we have
\begin{equation}
\Phi_B=-rg_B=-\frac{GM(r)}{r}\label{PhiBrgB},
\end{equation}
where $M(r)$ is the mass inside the sphere with radius $r$. In contrast, the Newtonian potential $\Phi_N$ is given by 
\begin{equation}
\Phi_N=\int_\infty^{r} g_B(r') dr'=\int_\infty^r \frac{GM(r')}{r'^2} dr'.
\end{equation}
The two potentials coincide when there is a point mass in the origin and no mass elsewhere. Note that $\Phi_B$ does not have any $a_0$ factor, because it is purely ``non-Verlinde gravitational'' quantity, even though Verlinde gravity uses it. Tolman-Oppenheimer-Volkoff calculated $\Phi_B$ in the presence of spherical symmetry, and obtained exactly (\ref{PhiBrgB}). However, to authors' knowledge, nobody suggested (\ref{nablaini}) which is our basis of calculation in the absence of spherical symmetry.   

Plugging $d=4$ and (\ref{SigmaDi}), we get
\begin{eqnarray}
\left(\frac{8\pi G}{a_0}\left(\frac{-2g_{Di}}{8\pi G}\right)\right)^2&=&\frac 23\nabla_i\left(\frac{\Phi_B}{a_0}n_i\right)\\
g_D^2&=&\frac{a_0}{6}\nabla_i (\Phi_B n_i).
\end{eqnarray}
By plugging (\ref{nablaini}), we get:
\begin{equation}
g_D^2=\frac{a_0}{6}\left((\nabla_i \Phi_B )n_i+2 g_B\right)\label{gD2}.
\end{equation}

Recalling the expression for $\vec{n}$ in (\ref{vecn}), we calculate its divergence,

\begin{equation}
\nabla\cdot \vec n = \frac{\nabla\cdot\vec g_B}{g_B}+\vec g_B\cdot\nabla\left(\frac{1}{g_B}\right) = -\frac{4\pi G\rho_B}{g_B}-\frac{\vec n}{g_B}\cdot \nabla g_B,
\label{nablagradient}
\end{equation}

\noindent
where we replaced $\nabla \cdot \vec{g}_B$ in the first term by using Gauss' law, i.e., $\nabla \cdot \vec{g}_B = -4 \pi G \rho_B$, where $\rho_B$ is the density of baryonic matter.

Finally, by replacing (\ref{nablagradient}) into (\ref{nablaini}), we get

\begin{equation}
\Phi_B=-\frac{2g_B^2}{4\pi G \rho_B+\vec n \cdot \nabla g_B}\label{PhiBmiddle}.
\end{equation}


As an aside, we want to mention that one can easily derive Tully-Fisher relation from (\ref{gD2}). Let's say the total mass of galaxy is $M_{tot}$. Then, plugging $g_B=GM_{tot}/r^2$ into (\ref{PhiBrgB}) (in which case we have $\Phi_B=\Phi_N$ as mentioned), and then into (\ref{gD2}) yields
\begin{equation}
(\nabla_i \Phi_B)n_i=-\frac{GM_{tot}}{r^2}=-g_B,\qquad
g_D^2=\frac{a_0}{6} g_B,
\end{equation}
which is exactly Tully-Fisher relation upon the identification $a_0/6$ with Milgrom's constant, which agrees with the one obtained from Hubble's constant within 10\%.

\section{Our calculation strategy}\label{section:strategy}
In this article, we are considering Verlinde gravity in the Solar System, where the typical gravitational acceleration is much bigger than $a_M$. Therefore, to a very good approximation, (\ref{gB2gD2}) can be re-expressed as
\begin{equation}
g=g_B+\frac{g_D^2}{2g_B}\label{gBgD2/2gB}
\end{equation}

Given this, how would the consideration of Verlinde gravity change the orbit of planets? At first glance, it might seem that, in order to calculate the Sun's gravitational attraction of a planet, we compute $g_B$ (i.e., the Sun's Newtonian attraction of the planet) and $g_D$ from quantities that can be obtained from $\vec g_B$, such as $\nabla g_B$, with $\rho_B$ as an additional input. 

However, this is not the case. When we calculate the (Newtonian or relativistic) gravitational attraction of the Sun towards a planet, we ignore the planet's own gravitational field. A planet receives net zero gravitational force from its own gravitational field. The ground near the North Pole receives the gravitational force toward the Earth's center, which is cancelled by the ground near the South Pole, if we see the Earth as a whole. However, the case is not so when we consider Verlinde gravity.

\begin{figure}[ht!]
\centering\includegraphics[height=40mm]{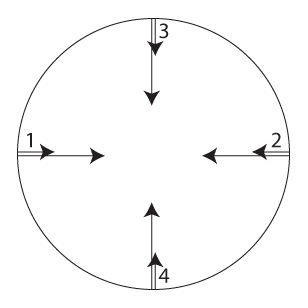}
\caption{Schematic representation of Earth's gravity on different points (1, 2, 3, 4). Long arrows represent $\vec g_B$, while short arrows represent $\vec g_D$. The figure and the length of the arrows are not to scale.}
\label{figure:gravity}
\end{figure}

Figure \ref{figure:gravity} displays a schematic representation of the Earth, together with some gravitational vectors, depicted as arrows. The long arrows denote $\vec g_B$ and the small arrows denote $g_D^2/(2g_B)$. It is worth noting that the picture is only representative and the length of the arrows are not to scale. If $\vec g_B$ at every point on the Earth is added, we will get the net gravitational attraction of the Sun $GM_S/R_S^2$, and we can forget about planet's own gravitational field. However, it is not so with $g_D^2/(2g_B)$. The vectors $\vec g_B$s at points 1, 2, 3, 4 shown in the Figure are not the same because of the gravitational attraction of the Sun. As an example, if the Sun is positioned far away to the left side of the planet, the long arrow at 1 will be shorter than the long arrow at 2, because the Sun's gravitational field is directed toward the left. This differences on the values of $g_B$ is responsible for the difference in $g_D^2/(2g_B)$. When $g_D^2/(2g_B)$ is added for all the interior points of the planet by integration (of course, assuming a vector sum by giving $g_D^2/(2g_B)$ the direction of $\vec g_B$), we get an additional non-zero pull towards the Sun. Later, we will calculate this pull, which we call $\vec g_{Vnet}$. It is also important to note that the obtained value of $\vec g_{Vnet}$ is very different from (actually much smaller than) $g_D^2/(2g_B)$ without considering the planet's own gravitational field. To repeat, the cancelation effect between the small arrows at 1 and 2 makes the effect of Verlinde gravity on planets much smaller than the one without such a consideration (i.e., without such a cancelation).

\section{Verlinde gravity on the Earth and other planets}\label{section:verlinde_on_planets}.

In this section we will calculate the value of Verlinde gravity for different planets. First, we start computing the value for the Earth as a starting point. Then, by switching the variable, we can easily calculate the Verlinde gravity on other planets. 

To calculate $g_D$, we need to calculate $g_B$ first, which is given by

\begin{equation}
\vec g_B=\vec g_E+\vec g_S.
\end{equation}
Here, $\vec g_E$ is the gravity due to the Earth, and $\vec g_S$ is the gravity due to the Sun. Of course, there are additional terms due to the gravity of other planets, but we will ignore them for the moment to simplify our analysis and come back to it again at the end of the section. 

Now, to calculate $g_B$ we do the following,
\begin{eqnarray}
g_B^2&=&(\vec g_E+\vec g_S)\cdot(\vec g_E+\vec g_S)\nonumber\\
&\approx&g_E^2-2g_E g_S\cos\theta_S\nonumber\\
g_B&\approx&g_E+\Delta g_B,
\end{eqnarray}
where
\begin{equation}
g_E=\frac{GM(r)}{r^2},\qquad\vec g_E=-g_E\hat r_E,\qquad \Delta g_B\equiv -g_S \cos\theta_S, \qquad\cos \theta_S=\hat r_E\cdot \hat r_S\label{expression}.
\end{equation}
In this expression, $\hat r_S$ is the unit vector pointing from the center of the Earth to the Sun.

In our case, (\ref{gD2}) and (\ref{PhiBmiddle}) can be re-expressed as follow:
\begin{equation}
g_D^2=\frac{a_0}{6}(2g_B-\partial_r \Phi_B)\label{gD2again},
\end{equation} 
with
\begin{equation}
\Phi_B=-\frac{2g_B^2}{4\pi G \rho_B-\partial_r g_B}.
\end{equation}
By using the definition of $M(r)$ and $g_S/R_S\ll g_E/R_E$, where $R_S$ is the distance to the Sun and $R_E$ is Earth's radius, i.e.,
\begin{equation}
4\pi r^2 \rho_B=\frac{\partial M(r)}{\partial r},\qquad \partial_r g_S\ll\partial_r g_E \label{rhoB}
\end{equation}
we get
\begin{eqnarray}
\partial_r g_B\approx \partial_r g_E=\partial_r\left(\frac{GM(r)}{r^2}\right)\\
=-\frac{2GM(r)}{r^3}+4\pi G \rho_B.\label{partialrgE}
\end{eqnarray}
Thus,
\begin{equation}
\Phi_B=-\frac{2 g_B^2}{2g_B/r}=-rg_B,\label{PhiB=-rgB}
\end{equation}
which implies
\begin{equation}
g_D^2=\frac{a_0}{6}(3g_B+r\partial_r g_B)\approx\frac{a_0}{6}(3g_B+r\partial_r g_E).
\end{equation}
Therefore, (\ref{gBgD2/2gB}) can be re-written as
\begin{equation}
g=g_B+\frac{a_0}{4}+\frac{a_0}{12 g_B}r\partial_r g_E
\end{equation}

Now, we need to integrate this for the whole interior of the Earth. Apparently, to obtain the net Verlinde gravity effect, we need to consider the last term only, because the first term will yield $g_S$, the Newtonian gravitational pull of the Sun, and the second term cancels out due to the spherical symmetry.

To calculate the third term, notice that
\begin{equation}
\Delta\left(\frac{a_0}{12 g_B}\right)=-\frac{a_0}{12 g_E^2} \Delta g_B.
\end{equation}
Thus,
\begin{equation}
M_E \vec g_{Vnet}=-\int \rho_B r^2 \sin \theta_S d\phi_S d\theta_S dr\left(\frac{a_0}{12 g_E^2}\Delta g_B r \partial_r g_E\cos\theta_S\right)\hat r_S,
\end{equation}
where $\vec g_{Vnet}$ is the additional pull toward the Sun due to Verlinde gravity, as we explained in Section \ref{section:strategy}. The factor $\cos\theta_S$ is due to the fact that only the component of gravity parallel to $\hat r_S$ survives due to the spherical symmetry. For example, in Fig. \ref{figure:gravity}, this factor is 0 for the arrows at 3 and 4 of Figure \ref{figure:gravity}, $-1$ for the arrow at 1, and 1 for the arrow at 2. By plugging $\Delta g_B$ in (\ref{expression}), we get
\begin{equation}
M_E \vec g_{Vnet}=\frac 43 \pi \left(\frac{a_0}{12}\right)\left(g_S\hat r_S\right) \left(\int_0^{R_E} \frac{\rho_B r^3}{g_E^2}\partial_r g_E dr\right)\label{MEgVSun}.
\end{equation}
Summarizing, the total force $\vec F$ on the Earth is given by
\begin{equation}
\vec F=M_E(\vec g_S+\vec g_{Vnet})\label{vecF}.
\end{equation}

Notice that (\ref{MEgVSun}) implies that
\begin{equation}
\vec g_{Vnet}=\lambda_E (g_s\hat r_S)=\lambda_E\vec g_S\label{VSun},
\end{equation}
where
\begin{equation}
\lambda_E\equiv\frac{\pi a_0}{9M_E}\int_0^{R_E} \frac{\rho_B r^3}{g_E^2}\partial_r g_E dr\label{lambdaE}.
\end{equation}

If we consider the effect of the Moon, we have the same expression as the above one, except that $\vec g_S$ is replaced by $\vec g_M$. If we consider all the effects, including those of other planets, which we denote by $\cdots$, we have
\begin{equation}
\vec g_{Vnet}=\lambda_E(\vec g_S+\vec g_M +\cdots).
\end{equation}
If we write the total external Newtonian gravity by $g_{Newton}$ as follows
\begin{equation}
\vec g_{Newton}=\vec g_S+\vec g_M+\cdots,
\end{equation}
(\ref{vecF}) can be updated to 
\begin{equation}
\vec F=M_E(\vec g_{Newton}+\vec g_{Vnet})=\left(1+\lambda_E\right)M_E \vec g_{Newton}
\end{equation}

If we write
\begin{equation}
\vec F=M_E \vec a=M_{E\mathrm{grav}} \vec g_{Newton},
\end{equation}
where $M_E$ is the inertial mass of the Earth, and $M_{E\mathrm{grav}}$ is the gravitational mass of the Earth, we see that our equation can be interpreted as the breakdown of equivalence principle for the Earth and other planets, i.e., $M_E\neq M_{E\mathrm{grav}}$. In particular, we have
\begin{equation}
\vec a=\left(1+\lambda_E\right)\vec g_{Newton}\label{veca}
\end{equation}
and
\begin{equation}
\frac{M_{E\mathrm{grav}}}{M_E}=1+\lambda_E.
\end{equation}

In the following section, we will show the results of the calculations of $\lambda$ for the different planets of our Solar System.

\section{The calculation of $\lambda$ for the planets and the Moon}\label{section:lambda_calculation}
We present in this Section the results of the calculation of $\lambda$ for the different planets of our Solar System and for the Moon. First, in the next section, we will compute $\lambda$ for the terrestrial planets and the Moon. Then, in the following section, we will compute those for the giant planets.

\subsection{Terrestrial planets and the Moon}
For simplification reasons to our model, we assume that terrestrial planets are comprised by a core and a mantle, each with a constant density. In reality, the density within the core and the mantle depends on its depth.  
We denote the planet's mass and radius by $M$ and $R$, respectively. We then define the baryonic density as

\begin{equation}
\rho_B=
\left\{
 \begin{array}{ll}
  \rho_c~~~&\mathrm{,}~ 0<r<R_c\\
  \rho_m~~&\mathrm{,}~R_c<r<R, 
 \end{array}
 \right.
\end{equation}

\noindent
where $\rho_c$ and $\rho_m$ are the planet's core and mantle density, respectively, and $R_c$ is the core radius.

The data considered as physical parameters for the terrestrial planets were taken from \cite{Mercury, Venus, Mars} and are summarized on Table \ref{table:terrestrial}.

\begin{table}[ht!]
	\begin{center}
		\begin{tabular}{|l|r|r|r|r|}
			\hline
			Planet & $R_c$ (km) & $R$ (km) & $\rho_c$ (kg/m$^3$) & $\rho_m$ (kg/m$^3$)\\ 
			\hline
			Mercury & 2002 & 2439 & 7245 & 3182   \\
			Venus & 3228 & 6052 & 10600 & 4300     \\
			Earth & 3486 & 6378 & 10700 & 4500   \\
			Mars & 1700 & 3389 & 6533 & 3554  \\
			\hline
		\end{tabular}
		\caption{Physical parameters for the terrestrial planets. $R_c$ is the planet's core radius, $R$ its radius, and $\rho_c$ and $\rho_m$ are the core and mantle densities, respectively.}
		\label{table:terrestrial}
	\end{center}
\end{table}

To compute the value of $\lambda$ for the Moon, we followed \cite{Weber2011}. From there, we modelled the Moon as an inner core, an outer core, and a mantle. With that consideration, the values adopted were the following:

\begin{equation}
 \rho_{Moon}(kg/m^3) = 
 \left\{\begin{array}{rl}
         8 \times 10^3 & ,0 < r < 240 \rm{ km}, \\
         5.1 \times 10^3 &  ,240 \rm{ km} < r < 330 \rm{ km}, \\
         3.3249 \times 10^3 &  ,330 \rm{ km} < r < 1737.4 \rm{ km},
        \end{array}\right.
\end{equation}

Next we will consider the case of giant planets.

\subsection{Giant Planets}\label{section:giant_planets}
Giant planets have an envelope and a solid core. Unlike terrestrial planets, the envelope comprises most of their mass. Because of this, we must consider the following expression:
\begin{equation}
\frac{dP}{dr}=-\rho_B g_E=-\rho \frac{Gm}{r^2},
\end{equation}
which implies
\begin{equation}
m=-\frac{r^2}{\rho_B G}\frac{dP}{dr}.
\end{equation}
If we plug this into Eq. (\ref{rhoB}), we obtain
\begin{equation}
\frac{d}{dr}\left(\frac{r^2}{\rho_B}\frac{dP}{dr}\right)=-4\pi G \rho_B r^2.
\end{equation}
Thus, if we know the relation between $P$ and $\rho$, we can solve the above second order differential equation. Such relations are available for Jupiter and Saturn in \cite{Relation}, and for Juptier in \cite{Jupiterrelation}. For our calculations, we used the density profile of Jupiter available in \cite{Jupiterdensity}, and the density profile of Saturn available in \cite{Saturndensity}.\\

For the cases of Uranus and Neptune, we considered two scenarios. 

\begin{itemize}
	\item First scenario: the core radius of these planets are one fifth of their radii (i.e., $R_c=0.2 R$). For their density, we used the profile described in \cite{Uranusandneptune}. In particular, Ravit Helled kindly provided the following formula \cite{Ravit}. The unit of the densities is in kg/m$^3$:
\end{itemize}

Uranus: $\rho_c=8231.99$,
\begin{equation}
\rho_{\mathrm{envelope}}  = 4039.37 + 20.7051 \beta^2 + 3.78416 \beta^3 - 38675.2 \beta^4 + 53208.8 \beta^5 -  18597.6 \beta^6
\end{equation}

Neptune: $\rho_c=10598.5$,
\begin{equation}
\rho_{\mathrm{envelope}} =  4614.26 - 23.9265 \beta^2 - 4.37289 \beta^3 - 29734.9 \beta^4 + 32288.5 \beta^5 - 7139.54 \beta^6,
\end{equation}
where $\beta$ is the normalized radius $\beta=r/R$, with $R$ the planet's radius. 

\begin{itemize}
	\item Second scenario: No core. For this scenario we considered that the giant planets don't have a solid core. As in the case of the first scenario, Ravit Helled kindly provided the following formula \cite{Ravit}. The unit of the densities is in kg/m$^3$:
\end{itemize}

Uranus:
\begin{equation}
\rho_{\mathrm{envelope}}=4424.91 - 49167.8 \beta^4 + 71977.3 \beta^5 - 27234. \beta^6
\end{equation}

Neptune:
\begin{equation}
\rho_{\mathrm{envelope}}=5145. - 44515.5 \beta^4 + 58559.1 \beta^5 - 19188.1 \beta^6.
\end{equation}

\subsection{Results}
For the sake of precision, there is no need to compute the value of $\lambda$ with more than two significant digits, as obtained in Section \ref{section:giant_planets}. The reason for this is that the current measurement of the Hubble's constant has only two significant digits of precision. For this work, we adopted the value for the Hubble's constant of $H_0=70$ km/s/Mpc. In Table \ref{table:lambda}, we present the results of the computation of $\lambda$ for the terrestrial and giant planets and for the Moon. In the table, $\lambda_0$ corresponds to $\lambda$ in the constant density model of planets. The value of $\lambda_0$ is given by
\begin{equation}
\lambda_0=\frac{a_0}{24 g_E(R_E)},
\end{equation}
where $g_E(R_E)$ is the surface gravity of the Earth (or, correspondingly, each planet). In Appendix \ref{Appendix}, we prove that $\lambda$ is always smaller than $\lambda_0$.

\begin{table}[ht!]
	\centering
	\begin{tabular}{|l|c|c|c|}
		\hline
		 & $\lambda_0$ ($\times10^{-12}$) & $\lambda$ ($\times10^{-12}$) & $\lambda$ ($\times10^{-12}$)\\ 
		\hline
		Mercury & 7.66& 4.59 &\\
		Venus & 3.20& 1.01 & \\
		Earth & 2.89 & 0.96 &\\
		Mars & 7.60& 3.80 &\\
		Jupiter &1.14 & -0.10 & -0.17 \\
		Saturn & 2.71  & -0.19 &\\
		Uranus& 3.26 & -0.03 & 0.33\\
		Neptune & 2.53& 0.00 & 0.32\\
		Moon & 17.47 & 14.65&\\
		\hline
	\end{tabular}
	\caption{Values of $\lambda_0$ and $\lambda$ for the planets in the Solar System and the Moon. The results are expressed in unit of $\times 10^{-12}$. $\lambda_0$ is the value of $\lambda$ in the constant density model of planets. For the calculation of $\lambda_0$, we used the surface gravity available in \cite{surfacegravity}.}
	\label{table:lambda}
\end{table}

In Table \ref{table:lambda}, we see Jupiter, Uranus and Neptune have two values of $\lambda$ computed. In the case of Jupiter, both values correspond to the calculation of $\lambda$ using two different core densities, 10 g/cm$^{3}$ (third column) and 100 g/cm$^{3}$ (fourth column), as described in \cite{Jupiterdensity}. For Uranus and Neptune, the two values were computed using the two considerations described in Section \ref{section:giant_planets}, i.e: with a core and an envelope (third column) and considering a planet without a core (fourth column).

\section{Comparison with data}

In Section \ref{section:lambda_calculation}, we defined and computed the value of $\lambda$ for the planets in our Solar System. In this section, we will analyze how the inclusion of Verlinde gravity affects different physical phenomena.

\subsection{The perihelion precession}
The first effect that we analyze is the precession of the perihelion of a planet due to Verlinde gravity. We found out that the effects of the precession of planets caused by Verlinde gravity is small. Moreover, the value of this effect is smaller than the error in the observations. Let's recall that the perihelion precession is present only when the total gravitational force on the planet is not proportional to the inverse square of the distance from the Sun. In other words, it only depends on the gravitational attraction of other planets, and the deviation from the inverse square law from Sun's gravitational force. This concept can be written in the following way:$\footnote{More precisely, the first term is given by $(3\pi/2)(m/M)(r/R)^3$ in the lowest approximation. In this expression, $m$ is the mass of the other planet, $M$ is the mass of the Sun, and $r$ is the size of the orbit. See \cite{Newtonianprecession}.}$

\begin{equation}
\mathrm{precession~of~perihelion}\propto\frac{\mathrm{gravitational~attraction~of~other~planets}}{\mathrm{gravitational~attraction~of~the~Sun}}+\mathrm{relativistic~effects}.\label{perihelion}
\end{equation}

Let's see how these two terms change when we consider Verlinde gravity. In the first term on the right side of the expression, both the numerator and the denominator are multiplied by the factor $(1+\lambda)$. Thus, as this factor is cancelled out, the first term becomes unaltered by the implementation of Verlinde gravity, at least up to order $\mathcal O(\lambda)$. The expression for the second term is given by

\begin{equation}
\Delta \theta_{\mathrm{rel}}=\frac{24\pi^3 a^2}{c^2T^2(1-e^2)}=6\pi \frac{m^2}{l^2},
\label{eq:relativistic}
\end{equation}
where $l=r^2\dot{\theta}$ is the angular momentum divided by the planet's mass. As we include Verlinde gravity, the planet's mass changes, from $m$ to $m(1+\lambda)$. Since the value for $m$ is squared, we will have a factor $(1+\lambda)^2$. By developing the power and discarding the quadratic term, we obtain that the expression of $\Delta \theta_{\mathrm{rel}}$ in Eq. (\ref{eq:relativistic}) will be proportional to $2\lambda$.

In the case of Mercury, the first term in (\ref{perihelion}) has a value of 532.3 arcsec/century, while the second term is 43.0 arcsec/century. Thus, the observed precession of the perihelion is 574.10$\pm$0.65 arcsec/century \cite{Mercuryprecession}. Please note that 43.0 arcsec/century multiplied by $2\lambda$ is 6.5$\times 10^{-10}$ arcsec/century, which is much smaller than the error in the observation, which is 0.65 arcsec/century. As we noted at the beginning of this section, the Verlinde gravity effect on the precession of the perihelion is too small to notice.

\subsection{Masses of the Sun and planets}\label{subsection:massesofthesun} 
At the end of Section \ref{section:verlinde_on_planets}, we interpreted the Verlinde effect as the breakdown of equivalence principle, i.e., the difference between the inertial mass and the gravitational mass. Another equally valid interpretation is that the masses of the Sun and other planets, which a planet ``feels'', is multiplied by the factor $(1+\lambda)$. However, masses of the Sun and planets are known much less precisely than $\mathcal O(\lambda)$ for the planets. Even the most precisely known mass is the one of the Sun, which is given by \cite{Astrodynamicsconstant},
\begin{equation}
GM_S=(1.32712440018\times 10^{20}\pm 8\times 10^9) \mathrm{m^3/s^2}.
\end{equation} 
Thus, the relative error is around $6\times 10^{-11}$, which is, at least, 10 times larger than $\lambda$s.

\subsection{Observational uncertainties in the position of planets}
The last subsection is enough to prove that Verlinde gravity effects on the orbits of planets are too small to be noticeable, but let us add in this section that observational uncertainties in the position of planets are too big to notice the Verlinde gravity effect either. To obtain the rough estimate in the uncertainties in $GM_S$, consider Kepler's 3rd law.

\begin{equation}
GM_S\approx r^3\omega^2+\mathrm{effects~of~other~planets}.
\end{equation}
Considering that $r^3\omega^2$ is much larger than the effects of other planets, we have
\begin{equation}
\frac{\delta(GM_S)}{GM_S}\approx 3\frac{\delta r}{r}+2\frac{\delta \omega}{\omega}.
\end{equation}

The distance between the Earth and the Sun is estimated by the distance between the Earth and other planets, and is known about by 1 meter uncertainty \cite{Folkner}. Given that the distance between the Earth and the Sun is about $1.5\times 10^{11}$ meters, we obtain
\begin{equation}
3\frac{\delta r}{r}\approx 2\times 10^{-11},
\end{equation}
which is much larger than $\lambda$s.

The uncertainties in the shape of the orbits of Jupiter and Saturn are about 10 meter \cite{Folkner}, which yields
\begin{equation}
 3\frac{\delta r}{r}\approx
 \left\{\begin{array}{l}
  4\times 10^{-11}~(\mathrm{for~Jupiter}),\\
  2\times 10^{-11}~(\mathrm{for~Saturn}),
 \end{array}
 \right.
\end{equation}
which are again much larger than $\lambda$s. We do not need to consider Neptune and Uranus, because their uncertainties are much larger as there is no series of spacecraft radio range measurements available from the other planets \cite{Folkner}.

Therefore, the uncertainties in $r$ are not small enough to notice the Verlinde gravity effect. Nevertheless, let's consider the uncertainties in $\omega$ as well. The uncertainties in $\omega$ come mainly from our ability to measure the orbit planes with respect to extra-galactic quasars that astronomers use to define the coordinate system \cite{Folkner}. We can use VLBI observations of spacecraft to measure this to about 0.25 milli-arcseconds, or 1.25 nanoradians \cite{Folkner}. This is both for inner planets in the past and the present and for outer planets in the present. Considering that the Earth orbited around the Sun about 300 radians after the first spacecraft was sent, we can estimate
\begin{equation}
2\frac{\delta \omega}{\omega}\approx 8\times 10^{-12},
\end{equation}  
which is larger than $\lambda$.

\subsection{The Mass of the Earth}
Just as in (\ref{subsection:massesofthesun}), the Moon feels that the mass of the Earth is multiplied by $(1+\lambda)$. The observed mass of the Earth is given by \cite{Astrodynamicsconstant}
\begin{equation}
GM_E=(3.986004418\times 10^{14} \pm 8\times 10^5)~\mathrm{m^3/s^2}.
\end{equation}
Thus, the relative error is approximately $2\times 10^{-9}$, which is about 100 times larger than $\lambda$ of the Moon.

\subsection{Observational uncertainties in the position of the Moon}
Currently, the lunar orbit is known to submeter accuracy \cite{Folkner}. Considering that the distance between the Earth and the Moon is about 380,000 km, we have (assuming $\delta r\approx 0.2\sim 0.5$ m)
\begin{equation}
3\frac{\delta r}{r}\approx (2\sim 4)\times 10^{-9},
\end{equation}
which is about 100$\sim$200 times larger than $\lambda$.

\section{Conclusions}
In this work, we computed the values for the Verlinde gravity's parameter, $\lambda$, for the different planets and the Moon in our Solar System. There, we found that the error bars in the mass of the Sun and the Earth are much larger than the effects of Verlinde gravity on the orbits of planets and the Moon. Furthermore, we found that it's not even possible to observe such effects, if there is no new development to drastically reduce the observation errors in the distance determination of the orbits.

As we mention at the beginning of this work, there are people who claim that Verlinde gravity effects on the planets should be great. In light of the results obtained in this work, we conclude that those asseverations are incorrect. The reason for this discrepancy is that either they don't consider the effect of planets own gravitation nor use our Verlinde gravity expression (\ref{gD2}), which is valid beyond spherical symmetry. For example, in \cite{MilgromSanders}, when analyzing the Verlinde gravity effects on the planets, the authors considered the following expression from \cite{Verlinde}
\begin{equation}
g_D=\sqrt{a_M g_B}=\sqrt{\frac{a_0}{6} g_B},
\end{equation}
which is only valid when there is a spherical symmetry. This is a special case of (\ref{gD2}), when 
\begin{equation}
g_B=\frac{GM_S}{R_S^2},\quad -\Phi_B=\frac{GM_S}{R_S}, \quad \rho_B=0, \quad \vec n\cdot\nabla g_B=\frac{2GM_S}{R_S^3}.
\end{equation}

As we have seen in Section \ref{section:verlinde_on_planets}, none of the above expressions are correct when applied to Verlinde gravitation on the orbits of planets. In the leading order, $g_B$ is given by planet's own Newtonian gravitation, instead of that of the Sun. $\Phi_B$ is also determined by each planet's own gravitation instead of the Sun's. $\rho_B$ is not certainly zero, as planets have non-zero mass density. For the derivative of Newtonian gravitational field, planet's own one dominates again.

In Newtonian gravity, we do not need to consider the effects of planets' own gravitation when calculating their motion, but when Verlinde gravity comes in, we need to consider them as it can clearly be seen from our demonstration in this paper; this consideration makes Verlinde gravity effects on planet's motion much smaller. 

It is worth to mention that, when we actually perform Verlinde gravity experiments on the Earth, such as the ones described in \cite{VerlindeYoon}, it is not necessary to consider the test particle's own gravity as it is negligible; when the test particle is just a small metal, its own gravity never dominates, but Earth's gravitation does. The amplitude and the direction of the combined gravity are almost the ones for the gravity of the Earth. However, when the test particle is a planet, its own gravity dominates; the amplitude and the direction of the combined gravity are very close to the ones of the planet's own gravity, not the ones of the Sun's gravity.

\begin{center}
	{\Large\bf Acknowledgments}
\end{center}
We thank Ravit Helled for providing the density profiles for Uranus and Neptune.
\appendix
\section{Proof that $\lambda$ is smaller than $\lambda_0$}\label{Appendix}

The relevant integration we need to consider is the following expression in (\ref{lambdaE}), which we call $b$.
\begin{eqnarray}
b=\int_0^{R_E} \frac{\rho_B(r) r^3}{g_E^2(r)}\partial_r g_E(r) dr
=\int_0^{R_E}\frac{\rho_B(r) r^3}{g_E^2(r)}(-2g_E(r)/r+4\pi G \rho_B(r)),
\end{eqnarray}
where we used Eq. (\ref{expression}). Now, if we define $\rho_M(r)$, the average density inside a sphere of radius $r$ as follows
\begin{equation}
\rho_M(r)=\frac{M(r)}{\frac 43 \pi r^3},
\end{equation}
we have
\begin{equation}
b=\int_0^{R_E} \frac{\rho_B(r)r}{\frac 43\pi G \rho_M^2(r)}(3\rho_B(r)-2\rho_M(r))dr.
\end{equation}
Notice that in the constant density model, we have the following:
\begin{equation}
b_0=\int_{0}^{R_E}\frac{r}{\frac 43\pi G}dr.
\end{equation}

Let's see in which cases we have $b<b_0$ (i.e., $\lambda<\lambda_0$). $b<b_0$ is satisfied if their respective integrands satisfy the same inequality, which implies
\begin{equation}
\frac{\rho_B(r)}{\rho_M^2(r)} (3\rho_B(r)-2\rho_M(r))<1,
\end{equation} 
which is equivalent to
\begin{equation}
\frac{3\rho_B}{\rho_M}-2<\frac{\rho_M}{\rho_B}.
\end{equation}
It is easy to check that the above condition is satisfied if
\begin{equation}
\rho_M(r)>\rho_B(r),
\end{equation}
which is always satisfied as $\rho_B(r)$ is always decreasing as $r$ is increasing. This completes the proof.

\end{document}